%% file: satellites_article.tex
\documentclass[twocolumn,10pt]{IEEEtran}
\usepackage{amsmath,amssymb,amsfonts}
\usepackage{algorithmic}
\usepackage{graphicx}
\usepackage{textcomp}
\usepackage{amsmath,amssymb,epsfig}
\usepackage{multicol}
\usepackage{varioref}
\usepackage{subfigure}
\usepackage{blkarray}
\usepackage{arydshln}
\usepackage{enumerate}
\usepackage{comment}
\usepackage{float}
\usepackage{color}
\usepackage{xfrac}
\usepackage{lipsum}
\usepackage{flushend}
\usepackage{mathtools}
\usepackage{cuted}
\usepackage{lscape}
\usepackage{hologo}
\usepackage[sort&compress,square,numbers]{natbib}
\def\BibTeX{{\rm B\kern-.05em{\sc i\kern-.025em b}\kern-.08em
		T\kern-.1667em\lower.7ex\hbox{E}\kern-.125emX}}
\input{./myshorts}

\graphicspath{{images/satellitesImages/}}
	\newif\ifpaper
\papertrue      
	\ifpaper
\newcommand{\TwoOneColumnAlternate}[2]{#1}
\else
\newcommand{\TwoOneColumnAlternate}[2]{#2}
\fi
\begin{document}
	
	\title{Asymptotic Performance of TDOA Estimation using Satellites}
     \author{Hodaya Halevi,~\IEEEmembership{Student Member,~IEEE,}
		Itsik Bergel,~\IEEEmembership{Senior Member,~IEEE}
		and Yair Noam,~\IEEEmembership{Member,~IEEE}%
		
		\thanks{H. Halevi, I. Bergel and Y. Noam are with the Faculty of engineering, Bar-Ilan University, Ramat-Gan, 5290002 Israel  (e-mail: hodaya.halevi.613@gmail.com; yair.noam@biu.ac.il). }
	}
	\maketitle

	\begin{abstract}
	We present novel lower bounds on the localization error using a network of satellites randomly deployed on a sphere around Earth. Our new analysis approach characterizes the localization performance by its asymptotic behavior as the number of satellites gets large while assuming a dense network. Using the law of large numbers, we derive closed-form expressions for the asymptotic Cram\'er Rao bound (CRB) from which we draw valuable insights. The resulting expressions depend solely on the network statistics and are not a function of a particular network configuration. We consider two types of estimators. The first uses the exact statistical model, and hence employs both timing and amplitude information. The second estimator ignores the amplitudes and hence uses only time difference of arrival (TDOA) information. The asymptotic CRB indicates that for practical system setup, a TDOA estimator approaches the performance of the ideal estimator. For both estimators, the localization accuracy improves as satellites get closer to Earth. The latter finding is essential in light of the proliferation of low-Earth-orbit (LEO) satellites and motivates a further study of localization-performance in such networks. Besides, we show that the vertical localization accuracy is lower than the horizontal accuracy and is also more sensitive to the receiver field-of-view.
	\end{abstract}
	
	\section{introduction}
	\input{./Chapters/sat_introduction}

	In this paper, we examine the case of numerous satellites and use the Law of Large Numbers to calculate the localization-accuracy asymptotic behavior, as a function of the number of satellites, for systems that employ either TDOA or TDOA+RSS localization. The analysis reveals new insights on satellite localization systems. In particular, for a given number of satellites, we show that localization accuracy is better as the satellites heights get lower, as long as we maintain sufficient coverage. We also show that the vertical localization accuracy is lower than the horizontal accuracy, and is also more sensitive to the receiver field of view. \par
	The remainder of this paper is as follows. In Section \ref{SEC:PORBLEMFORMULATION} we formulate the problem, including
	the signal model, assumptions about the system and the unknown parameters.
	In Section \ref{chap:derivation_rss}, we derive the asymptotic CRB for TDOA+RSS receiver localization in satellites. In Section \ref{chap:derivation_tdoa}, we derive the asymptotic performance of TDOA only receiver localization via a satellite network, and show that it is close to optimal in practical setups. 
	The section also includes valuable insights. Simulation results appear in Section \ref{chap:results}. Section \ref{chap:conclusions} includes a summary and conclusions. 
	\par
\textit{Notation}: Bold face  upper (lower) case letters denotes matrices (vectors), $[\Amat]_{i,j}$ denotes the $i,j$ entry of the matrix  $\Amat$ and $\mathbf{A}^{\top}$ is its transpose. For brevity, we use $\sum_{i}(\cdot)$ for $\sum_{i=1}^{N}(\cdot)$. 

\section{system model} \label{SEC:PORBLEMFORMULATION}
\input{./Chapters/sat_formulation_rss_v2}

\input{./Chapters/sat_acrb_derivation}


\section{numerical results}\label{chap:results}
\input{./Chapters/sat_numerical_results}


\section{Conclusions} \label{chap:conclusions}
\input{./Chapters/sat_conclusions}

\begin{appendices}
\input{./Chapters/sat_proofs}
	
\end{appendices}

\begin{small}
	\bibliography{BibSatellitesHodaya}
	\bibliographystyle{ieeetr}
\end{small}
\end{document}

%% file: myshorts.tex
\newcommand{\bbmat}{\begin{bmatrix}}
\newcommand{\ebmat}{\end{bmatrix}}

\floatstyle{ruled}%
\newfloat{floatequation}{thbp}{leq}
\floatname{floatequation}{}

\newtheorem{theorem}{Theorem}

\newtheorem{lemma}[theorem]{Lemma}

\newtheorem{claim}[theorem]{Claim}

\newtheorem{definition}{Definition}

\newcommand{\rh}{\mbox{${\hat r}$}}

\newcommand{\beq}{\begin{equation}}
\newcommand{\eeq}{\end{equation}}
\newcommand{\bea}{\begin{array}}
\newcommand{\ena}{\end{array}}
\newcommand{\bds}{\begin {itemize}}
\newcommand{\eds}{\end {itemize}}
\newcommand{\bdf}{\begin{definition}}
\newcommand{\blm}{\begin{lemma}}
\newcommand{\edf}{\end{definition}}
\newcommand{\elm}{\end{lemma}}
\newcommand{\bthm}{\begin{theorem}}
\newcommand{\ethm}{\end{theorem}}
\newcommand{\bprp}{\begin{prop}}
\newcommand{\eprp}{\end{prop}}
\newcommand{\bcl}{\begin{claim}}
\newcommand{\ecl}{\end{claim}}
\newcommand{\bcr}{\begin{coro}}
\newcommand{\ecr}{\end{coro}}
\newcommand{\bquest}{\begin{question}}
\newcommand{\equest}{\end{question}}

\def\bm#1{\mbox{\boldmath $#1$}}

\newcommand{\avec}{{\bf{a}}}

\newcommand{\rvec}{{\bf{r}}}

\newcommand{\onevec}{{\bf{1}}}

\def\xivec{{\mbox{\boldmath $\xi$}}}

\newcommand{\Amat}{{\bf{A}}}

\newcommand{\Jmat}{{\bf{J}}}

\def\gammavec{{\mbox{\boldmath $\gamma$}}}

\def\thetavec{{\mbox{\boldmath $\theta$}}}

\newcommand{\beqna}{\begin{eqnarray}}
\newcommand{\eeqna}{\end{eqnarray}}

%% file: Chapters/sat_introduction.tex
Localization using satellites has long been widespread but has attracted renewed interest due to the launching of rising numbers of satellites \cite{jaldehag2018evaluation,li2019leo,montenbruck2017multi}. 
Traditional global navigation satellite systems (GNSS) include the American GPS, the Russian GLONASS (e.g., \cite{leick1998glonass}), the Chinese BeiDou (e.g., \cite{sun2001working}) and the European Galileo (e.g., \cite{closas2009cramer}). The current trend in GNSS utilizes multiple GNSS types to improve localization accuracy (e.g., \cite{li2015accuracy}). This trend underscores the importance of studying localization accuracy with a large number of satellites. 
Furthermore, the recent launches of low earth orbit (LEO) communication satellites (such as OneWeb \cite{de2015virgin}) enables new alternatives for high precision localization through these satellites. Each of these constellations is expected to have thousands of satellites. This paper focuses on the characterization of localization accuracy with a large number of satellites. \par
	In any GNSS, the user computes its position from signals transmitted by a set of visible satellites. These systems are commonly used for security, communication, surveillance, reconnaissance, traffic monitoring and other applications (e.g., \cite{klemm2004applications,runge2008traffic}).\par
	State-of-the-art localization methods utilize time difference of arrival (TDOA) \cite{ouyang2017direct,sun2018solution,cao2019combined,knapp1976generalized,yang2011efficient,yang2009efficient,myref12, myref11, weiss2003accuracy, Shang2009Multiuser, nlos1tdoa,van2007bayesian,ho1993solution,bhatti2012development,cetin2014passive}, 
	frequency difference of arrival (FDOA) or both \cite{zhao2017joint,pattison2000sensitivity,yang2011efficient}. Localization information is also embedded in the received signal strength (RSS) \cite{ghannouchi2012accurate, zhu2011ekf, kazikli2019hybrid, catovic2004hybrid, catovic2004cramer} but is generally known (and proved below) to be negligible for satellite localization. This work analyzes the asymptotic performance of a system with a large number of satellites using TDOA or TDOA+RSS. Asymptotic performance with FDOA is left for future research.\par 
The Cram{\'e}r-Rao lower bound (CRB) is an essential tool for the performance analysis of localization systems.
	However, in many configurations, mainly when there are many sensors, the CRB expression is complicated and hard to interpret. Moreover, an expression that depends on a specific configuration is inconvenient since satellite-network geometry varies over time.\par  
	Incorporating randomness can lead to a more informative expression, which depends solely on a few system parameters. 
	For example, in ground based localization, Zheng et al. \cite{zheng2010closed} investigated the sensor-density effect on localization performance and bounded the expected CRB for RSS-based estimation (using a non-standard attenuation model in which the signal decayed exponentially with the squared distance). Shoari and A. Seyedi \cite{shoari2011target} also treated the CRB as a random variable depending on the sensor locations. They provided closed-form lower and upper bounds on the localization performance in the presence of Rayleigh fading, using the Law of Large Numbers. Bergel and Noam \cite{YnoamIbergelTOA} derived the expected CRB for Poisson point process sensor-locations, using closed-form expressions for the CRB that were derived by \cite{sieskul2009hybrid}. 
  \nocite{gierull2008ground}	\par

%% file: Chapters/sat_formulation_rss_v2.tex
We consider the localization of a single land terminal (LT) at an unknown location through a network of $N$ satellites. The satellites are uniformly distributed on a sphere of radius $R=r_{\mathrm{e}}+\tilde{h}$ around the Earth, where $\tilde{h}$ is the satellite height relative to Earth surface, and $r_{\mathrm{e}}$ is Earth radius.\footnote{In reality, satellite constellation design aims at maximizing coverage. Thus the probability of two satellites being near each other is typically smaller than in a uniformly distributed constellation. However, the satellite distribution is becoming more uniform as their number increases. Furthermore, in the case of multi-GNSS, the combined distribution is typically close to uniform.}
Without loss of generality, we assume that the LT is horizontally located at the true North Pole. That is, the LT is at an elevation $\delta$ above the north pole. We also denote the distance of the LT from the Earth center by $r=r_{\mathrm{e}}+\delta$ . We will assume throughout that $\delta\ll r_{\mathrm{e}}$ and $\delta\ll h$. Thus, in the text we will refer to $r$ also as the Earth radius. Yet, we emphasize that the exact meaning of $r$ (and the only way that it is used in the equations) is the LT distance from the Earth center. For mathematical simplicity, it is also convenient to work with the relative satellite height $h=\tilde h-\delta$ (thus, $R=r+h$ and again $h$ is approximately equal to $\tilde h$). 
 
\begin{figure}[t]
 	\centering
 	\includegraphics[width=\TwoOneColumnAlternate{0.9}{0.7}\columnwidth]{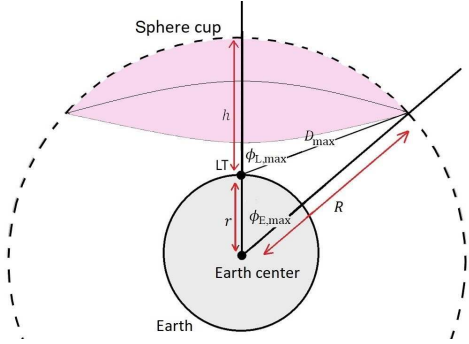}
 	\caption{System model: The satellites are randomly scattered around the Earth, and the LT can only receive from satellites located in the pink sphere cup defined by $\phi_{\rm {\rm L,max}}$.}
 	\label{fig:sys_mod} 
 \end{figure}
 
\par 
We use three coordinate-systems. The first is an Earth-centered, Earth-fixed (ECEF) \cite{clynch2006earth} cartesian right hinged system, used for reference and error calculation. The other two coordinate systems are spherical. One is centered at Earth center and is called the $\rm E$ coordinates. Here, the LT is located at the point $(r,0,90^\circ)$, that is, at radius $r$, azimuth $0$ and elevation $90^\circ$, respectively, whereas satellite-$m$ is at $(R,\theta_{{\rm E},m}, \phi_{{\rm E},m})$ where $R=r+h$. 
\par
The second spherical coordinate system is centered at the LT and is called the $\rm L$ coordinates. Note that azimuth is identical in both systems, $\theta_{{\rm L},m}=\theta_{{\rm E},m}$, whereas the elevation angles are not. The latter two satisfy
\begin{equation}
\sin\phi_{{\rm L},m}=\frac{R}{D_m}\sin\phi_{{\rm\tiny E},m},\quad \cos\phi_{{\rm \tiny L},m}=\frac{R\cos\phi_{{\rm\tiny E},m}-r}{D_m}
\label{coordinateTransformation}
\end{equation} 
where 
\begin{IEEEeqnarray}{rCl}
	D_m&=&\sqrt{R^2+r^2-2rR\cos\phi_{{\rm\tiny E},m}}\TwoOneColumnAlternate{\nonumber\\
	&=&}{=}\sqrt{R^2-r^2\sin^2\phi_{{\rm \tiny L},m}}-r\cos\phi_{{\rm \tiny L},m}
	\label{d_i_form}
\end{IEEEeqnarray} 
is the distance between the LT and satellite-$m$.
We assume the LT points upward (to space) and has a viewing angle $2\phi_{{\rm L,max}}$, $0<\phi_{{\rm L,max}}<\pi/2$, as depicted in Fig. \ref{fig:sys_mod}. We denote satellite $m$ location by $\xivec_{m}=[x_{m},y_{m},z_{m}]^\top$ and collect all these points in $\xivec_{1}^{N}=[\xivec_{1}\ldots \xivec_{N}]$. Finally, we denote the unknown parameter of interest (the LT position) by $\xivec=[x,y,z]^\top$.

The signal transmission begins at time $t = T_0$, unknown to the receiver. Typically, GNSS uses Code-Division Multiple Access (CDMA) to differentiate the signals from different satellites. For simplicity, we consider Time-Division Multiple Access (TDMA), in which the same signal is transmitted by different
satellites at different times. Note that the performance of TDMA and CDMA are nearly identical (e.g., \cite{botteron2004cramer}).\par 
We assume that different satellites transmit with a large enough time difference so there is no inter-satellite interference and the noise is independent while receiving these signals. Without contradicting the above, we also assume that the transmission time gap between satellites is small enough such that the movement of the LT during this time is negligible. The transmission time gap between satellites is known to the LT and hence has no effect on the performance. Thus, in the following we ignore this time gap, and simply assume that the LT can receive each of the satellites signal independently of the others. 

Satellite $m$ signal as received at the LT is given by \cite{YnoamIbergelTOA}
\begin{equation}
r_m(t)=D_m^{-1} s(t-\tau_m-T_0)\mathbf{1}_{[\phi_{{\rm L},m}<\phi_{{\rm L,max}}]}+v_m(t)
\label{rec_sig}
\end{equation}
where $t\in\mathbb{R}_+,m\in\mathbb{N}$, $s(t)$ is a known transmitted waveform, $\mathbf{1}_{[\cdot]}$ denotes the indicator function and $v_m(t)$ is a white Gaussian noise with spectral density $N_0/2$. The propagation time delay is $\tau_m = D_m/c$, where $c$ is the speed of light. 
In this paper we consider the estimation of $\xivec$ from the observations $\{r_{m}\}_{m=1}^{N}$, observed by the LT. 
We assume the following regularity conditions: $s(t)$  is continuous, differentiable,  equals $0$ for all $t\leq 0$ and 
there exists a non negative function $g(t)$, with $\int_{0}^{\infty}g(t)dt<\infty$, such that:
\begin{equation}
|s(t)|, \left|\frac{\partial s(t)}{\partial t}\right|\leq g(t)\qquad\forall t.
\end{equation}	
\par 
We will show that the performance can be characterized by the signal energy $E_{\rm s}\triangleq \int_{-\infty}^{\infty}s^2(t)dt$ and the effective bandwidth
\begin{equation}
W_e\triangleq\sqrt{\frac{\int_{-\infty}^{\infty}(\partial s(t)/\partial t)^2dt}{\int_{-\infty}^{\infty}s^2(t)dt}}
\end{equation}
Finally, we denote 
\begin{equation}
	\rho=E_s/N_0, \quad\eta=W_e^2/c^2
\end{equation}
 and denote the received signal energy if a satellite is exactly above the receiver by $E_{\rm s,max}=E_s/h^2$. Note that the latter is a reference energy and does not require an actual satellite at that location.  \par

%% file: Chapters/sat_acrb_derivation.tex
\section{Asymptotic Performance for TDOA+RSS}\label{chap:derivation_rss}

We now derive an asymptotic expression for the CRB as the number of satellites increases. We first derive the ${\rm CRB}(\xivec_1^N)$ on the estimation error of $\xivec$ as a function of the satellite locations, $\xivec_1^N$, and then evaluate its limiting behavior as $N$ increases. According to the Cram{\'e}r-Rao theorem \cite{rao1945information}, the maximum likelihood estimator performance converges to the CRB for high enough SNR. Note that during the derivation of ${\rm CRB}(\xivec_1^N)$, the satellites locations $\xivec_1^N$ are considered deterministic. 
\par

The estimation uses continuous-time signals, and hence, may exploit both TDOA and RSS information. 
The Fisher information matrix (FIM) for estimating $\gammavec$ is 
\begin{equation}
	\Jmat=\sum_i\Jmat_i\mathbf{1}_{\phi_{{\rm \tiny L},i}}
\end{equation}
where $\Jmat_i$ is the matrix that represents the contribution of sensor $i$ and given by \eqref{J3Drss}\TwoOneColumnAlternate{, at the top of the next page}{} (see \cite{myref12}),  $K_i=\frac{2\rho}{D_i^{4}}\Big(1+\eta D_i^2\Big) $, $\mathbf{1}_{\phi_{{\rm \tiny L},i}}=\mathbf{1}_{[\phi_{{\rm \tiny L},i}<\phi_{\rm L,max}]}$ and $L_i=\frac{2\rho\eta}{D_i^{2}}$. \par 
\begin{table*}[ht]
\fontsize{10}{12}{\begin{IEEEeqnarray}{rCl}
		&&\Jmat_i=
		\begin{pmatrix}
			K_i\sin^2\phi_{{\rm \tiny L},i}\cos^2\theta_i&
			K_i \sin^2\phi_{{\rm \tiny L},i}\cos\theta_i\sin\theta_i&
			K_i\sin\phi_{{\rm \tiny L},i}\cos\theta_i\cos\phi_{{\rm \tiny L},i}&
			-L_i\sin\phi_{{\rm \tiny L},i}\cos\theta_i\\
			K_i \sin^2\phi_{{\rm \tiny L},i}\cos\theta_i\sin\theta_i&
			K_i\sin^2\phi_{{\rm \tiny L},i}\sin^2\theta_i&
			K_i\sin\phi_{{\rm \tiny L},i}\sin\theta_i\cos\phi_{{\rm \tiny L},i}&
			-L_i\sin\phi_{{\rm \tiny L},i}\sin\theta_i\\
			K_i\sin\phi_{{\rm \tiny L},i}\cos\theta_i\cos\phi_{{\rm \tiny L},i}&
			K_i\sin\phi_{{\rm \tiny L},i}\sin\theta_i\cos\phi_{{\rm \tiny L},i}&
			K_i\cos^2\phi_{{\rm \tiny L},i}&
			-L_i\cos\phi_{{\rm \tiny L},i}\\
			-L_i\sin\phi_{{\rm \tiny L},i}\cos\theta_i&
			-L_i\sin\phi_{{\rm \tiny L},i}\sin\theta_i&
			-L_i\cos\phi_{{\rm \tiny L},i}&
			L_i
		\end{pmatrix}\label{J3Drss}
\end{IEEEeqnarray}}
\end{table*}
Let $\hat \xivec=[\hat x,\hat y,\hat z]^{\top}$ be an unbiased estimate of $\xivec$. Here, we consider three kinds of localization errors: the overall 3D error $e_{\rm xyz}=\Vert \hat \xivec-\xivec\Vert$, a horizontal error $e_{\rm xy}=\sqrt{(\hat x-x)^{2}+(\hat y-y)^{2}}$ and the altitude error  $e_{z}=|\hat z-z|$. The CRB on each error is the sum of the corresponding diagonal entries of the inverse FIM. Moreover, the first CRB is equal to the sum of the two others,
\begin{equation}
	\rm CRB_{xyz}(\xivec_1^N)=CRB_{xy}(\xivec_1^N)+CRB_z(\xivec_1^N)
	\label{CRBparts}
\end{equation}
 where ${\rm CRB_{xy}}(\xivec_1^N)=[\Jmat^{-1}]_{1,1}+[\Jmat^{-1}]_{2,2}$ and ${\rm CRB_{z}}(\xivec_1^N)=[\Jmat^{-1}]_{3,3}$.
Recalling that the satellite-locations $\xivec_1^N$ are random, it follows that the CRB is also random. Thus, we define the following limiting expressions:
\begin{IEEEeqnarray}{rCl}
	{\rm LCRB_{xy}}&=&\lim_{N\rightarrow\infty}N\cdot{\rm CRB_{xy}(\xivec_1^N)}\nonumber\\
	{\rm LCRB_{z}}&=&\lim_{N\rightarrow\infty}N\cdot{\rm CRB_{z}(\xivec_1^N)}	\nonumber\\
	{\rm LCRB}_{\rm xyz}&=&\lim_{N\rightarrow\infty}N\cdot{\rm CRB_{xyz}(\xivec_1^N)}\TwoOneColumnAlternate{\nonumber\\
	&=&}{=}{\rm LCRB_{xy}}+{\rm LCRB_z}.
	\label{lcrb_def}
\end{IEEEeqnarray}
where the limit is in the almost sure convergence sense. The following theorem expresses the first two quantities in closed form as a function of the system  parameters (see Fig. \ref{fig:sys_mod}).
\begin{theorem}
	For the system defined in Sec. \ref{SEC:PORBLEMFORMULATION}, ${\rm LCRB_{\rm xy}}$, ${\rm LCRB_{\rm z}}$ (\ref{lcrb_def}), are given by 
	\begin{IEEEeqnarray}{rCl}
		&&\hspace{-7mm}{\rm LCRB}_{\rm xy}=\frac{64Rr^3}{\rho}\Big[4 \left(2\eta\left(R^2+r^2\right)-1\right) \log \left(\frac{D_{\max}}{h}\right)\TwoOneColumnAlternate{\nonumber\\
		&&}{}+\frac{(2 \left(\eta(R^2-r^2)^2-2 (R^2+r^2) \right))(h^2-D_{\max}^2)}{D_{\max}^2h^2}\nonumber\\
		&&-4\eta r (h-D_{\max}\zeta)+\frac{\left(R^2-r^2\right)^2(h^4-D_{\max}^4)}{D_{\max}^4h^4}\Big]^{-1}
		\label{mcrb2drss}
	\end{IEEEeqnarray}

\begin{IEEEeqnarray}{rCl}
&&\hspace{-7mm}{\rm LCRB}_{\rm z}=\frac{16Rr^3}{\rho}\Bigg(2 \left(R^2-r^2\right) \left(\eta\left(R^2-r^2\right)-2\right)\TwoOneColumnAlternate{\nonumber\\
		&&\cdot }{}(h^{-2}-D_{\max}^{-2})-4 \left(2 \eta(R^2-r^2)-1\right) \log \left(\frac{D_{\max}}{h}\right)\nonumber\\
		&&+4 \eta r (h-D_{\max}\zeta)+\frac{\left(R^2-r^2\right)^2(D_{\max}^4-h^4)}{D_{\max}^4h^4}\TwoOneColumnAlternate{\nonumber\\
		&&}{}-\frac{16\eta\Big(D_{\max}(R+r\zeta)-(R^2-r^2)\Big)^2}{D_{\max}^2\log\Big(\frac{D_{\max}}{h}\Big)}\Bigg)^{-1}
		\label{mcrbzrss}
	\end{IEEEeqnarray}
where $\zeta=\cos\phi_{\rm L, max}$ and
\begin{equation}
	D_{\max}=\sqrt{R^2+r^2(\zeta^2-1)}-r\zeta.
	\label{DM}
\end{equation}
\label{THEOREM:LCRB_RSS}
\end{theorem}
\begin{IEEEproof}
	see Appendix \ref{PRTHEOREM:LCRB_RSS}.
\end{IEEEproof}
\par
While \eqref{mcrb2drss} and \eqref{mcrbzrss} seem complicated, they well characterize the achievable localization accuracy for a large number of satellites. Section \ref{chap:derivation_tdoa} provides further insights from these expressions. Before that, the following section characterizes the performance of estimators that only utilize timing data (TDOA). Such an approach is important because  (as proved bellow) amplitude data (RSS) contribution is negligible in satellite localization (with a reasonable transmission bandwidth).

\section{Asymptotic Performance for TDOA}\label{chap:derivation_tdoa}

\subsection{Estimation Based Solely on TDOA}
	An estimator that utilizes timing information while ignoring the signal strength information  is, by definition, sub-optimal. Thus, its performance cannot be characterized by the CRB of the problem. Instead, we need to consider the performance of a specific estimator. To this end we consider the maximum likelihood estimator in an alternative channel model, in which the signal strength is not a function of the receiver location, by replacing the signal in (\ref{rec_sig}) with 
	\begin{equation}
	\tilde{r}_{m}(t)=A_ms(t-\tau_m-T_0)\mathbf{1}_{[\phi_{{\rm \tiny L},m}<\phi_{\rm L,max}]}+v_m(t).
	\label{tdoamodel}
	\end{equation}
	Here, the overall unknown parameter vector is $\widetilde{\gammavec}=[\gammavec^{\top}, {\bf a}^{\top}]^{\top}$, where the amplitudes $\avec=[A_{1}\ldots, A_{N}]$ are a nuisance parameter-vector. 
	We denote the probability distribution induced by
	the alternative model as $\tilde P_{\text{\bf$\gamma$}}$ and the corresponding likelihood by $\tilde L(r_1\ldots r_N,\tilde\gammavec)$.
	In the sequel, we derive the ML estimator of $\tilde \gammavec$ from the distribution $\tilde P_{\text{\bf$\gamma$}}$. Note that the latter estimator considers the amplitudes, $\avec$, as a separate parameter vector, which is not a function $\{D_{m}\}_{m=1}^{N}$, hence not a function of the RSS.

\subsection{Mismodeling}
We now study the performance of the estimator derived while assuming that (\ref{tdoamodel}) holds while (\ref{rec_sig}) is the actual signal-model. Here, 
the expression for the CRB under (\ref{tdoamodel}), dubbed mismodeled CRB in this subsection, is not a lower bound on the mean square error (MSE) of every estimator. The problem of deriving estimators under a model that is not the actual probability distribution governing the observations is known as mismodeling. While the mismodeled CRB is not a general lower bound, it is a very useful tool if the performance of ML under mismodeling  approaches it asymptotically. A sufficient condition for such a convergence is the following:  
\begin{theorem}[\cite{white1982maximum}]\label{TheoremWhite1982}
	Considering an estimator that performs ML based on the likelihood $f(r(t);\bm{\theta})$ for estimating $\theta\in\Theta$ from the observation $r(t)$, where $g(r(t))$ be the likelihood function based on the actual distribution of $r(t)$. The ML-MSE of the considered estimator approaches the CRB derived under $f(r(t);\bm{\theta})$ if there exists $\bm{\theta}_{*}\in\Theta$, dubbed "legal" $\bm{\theta}_*$, such that $g(r(t))=f(r(t);\bm{\theta}_*)$.
\end{theorem}
\par We note that in the problem at hand, $A_{m*}=D_m^{-1}$ yields a ''legal'' parameter vector ($\thetavec_*$). Explicitly, considering Theorem \ref{TheoremWhite1982}, the actual distribution ($g(r(t))$ in the theorem) is defined by (\ref{rec_sig}) for a given value of $\gammavec$, say $\gammavec_{\rm a}$, while the parametric family is $\tilde L(r_1\ldots r_{N},\tilde \gammavec)$  ($f(r(t),{\bm \theta})$ in the theorem), the parametric family, $\tilde L(r_1,\ldots,L_N,\tilde {\bm \gamma})$ is obtained under the non-physical model in (\ref{tdoamodel}), but, it does include the actual likelihood. The latter is a special case of $\tilde L(\tilde \rvec,\tilde\gammavec)$ when substituting  $A_{m}=D_{m,*}^{-1}$ in  $\tilde \gammavec$, where $D_{m,*}$ is the actual distance between satellite-$m$ and the LT.\par
Thus, the ML performance are indeed characterized by deriving the CRB for the model in (\ref{tdoamodel}). While using the notation $\widetilde{CRB}$ for the CRB formula of the model (\ref{tdoamodel}), we remember that this is not an actual bound, and use it solely to characterize the ML performance.
In the next subsection, we show that this performance is very close to the true CRB (derived in Section \ref{chap:derivation_rss}) when $W_{\rm e}$ is large enough. Thus, this alternative model yields convenient, closed-form-expression for performance analysis.

\par 
The corresponding FIM for $\tilde{\gamma}$ estimation is
\begin{eqnarray}
\rm 	\tilde{\Jmat}_{extended}=\begin{pmatrix}
	 \tilde{\Jmat}&\Jmat_{\rm cros}\\
	\Jmat_{\rm cros}^H&\Jmat_{\rm amp}
	\end{pmatrix}
\end{eqnarray}
where $\Jmat_{\rm amp}=2\rho \textbf{I}_N$ is the FIM for estimating $\bf{a}$ and $\tilde{\Jmat}$ is the FIM for the estimation of $\bf{\gamma}$, which we present below. Now that the amplitudes are unknown nuisance parameters, we begin with evaluating their effect on the localization error. In Appendix \ref{chap:appendixA}, we show no coupling between the estimation of $\gammavec$  and the estimation of ${\bf a}$. Explicitly, $\Jmat_{\rm cros}$ is zero; hence $\rm \widetilde{CRB}$ for estimating $\gammavec$ is unchanged whether $\avec$ is estimated or known in advance. 
 Henceforth, we assume that the amplitudes are known and focus on the unknown $\gammavec$.
Similar to $\Jmat$ (see \eqref{J3Drss}), the FIM $\tilde{\Jmat}$ for estimating $\gammavec$ is 
\begin{equation}
	\tilde{\Jmat}=\sum_i\tilde{\Jmat}_i\mathbf{1}_{\phi_{{\rm \tiny L},i}}
\end{equation}
where $\tilde{\Jmat}_i$ is given by \eqref{J3D_t}\TwoOneColumnAlternate{, at the top of the next page}{}.\par 
\begin{table*}[ht]
\fontsize{10}{12}{\begin{IEEEeqnarray}{rCl}
		&&\hspace{-5mm}\tilde{\Jmat}_i=
		\begin{pmatrix}
			L_i \sin^2\phi_{{\rm \tiny L},i}\cos^2\theta_i&
			L_i \sin^2\phi_{{\rm \tiny L},i}\cos\theta_i\sin\theta_i&
			L_i \sin\phi_{{\rm \tiny L},i}\cos\theta_i\cos\phi_{{\rm \tiny L},i}&
			-L_i \sin\phi_{{\rm \tiny L},i}\cos\theta_i\\
			L_i \sin^2\phi_{{\rm \tiny L},i}\cos\theta_i\sin\theta_i&
			L_i \sin^2\phi_{{\rm \tiny L},i}\sin^2\theta_i&
			L_i \sin\phi_{{\rm \tiny L},i}\sin\theta_i\cos\phi_{{\rm \tiny L},i}&
			-L_i \sin\phi_{{\rm \tiny L},i}\sin\theta_i\\
			L_i \sin\phi_{{\rm \tiny L},i}\cos\theta_i\cos\phi_{{\rm \tiny L},i}&
			L_i \sin\phi_{{\rm \tiny L},i}\sin\theta_i\cos\phi_{{\rm \tiny L},i}&
			L_i \cos^2\phi_{{\rm \tiny L},i}&
			-L_i \cos\phi_{{\rm \tiny L},i}\\
			-L_i \sin\phi_{{\rm \tiny L},i}\cos\theta_i&
			-L_i \sin\phi_{{\rm \tiny L},i}\sin\theta_i&
			-L_i \cos\phi_{{\rm \tiny L},i}&
			L_i
		\end{pmatrix}\label{J3D_t}
\end{IEEEeqnarray}}
\end{table*}
Similar to the RSS-TDOA case, we consider three kinds of estimation errors:
\begin{equation}
\rm \widetilde{CRB}_{xyz}(\xivec_1^N)=\widetilde{CRB}_{xy}(\xivec_1^N)+\widetilde{CRB}_z(\xivec_1^N)
\label{til_CRBparts}
\end{equation}
where ${\rm \widetilde{CRB}_{xy}}(\xivec_1^N)=[\tilde{\Jmat}^{-1}]_{1,1}+[\tilde{\Jmat}^{-1}]_{2,2}$ and ${\rm \widetilde{CRB}_{z}}(\xivec_1^N)=[\tilde{\Jmat}^{-1}]_{3,3}$.
\subsection{Asymptotic Performance Analysis}
As in \eqref{lcrb_def}, we consider
\begin{IEEEeqnarray}{rCl}
	{\rm \widetilde{LCRB}_{xyz}}\TwoOneColumnAlternate{&=&}{=}\lim_{N\rightarrow\infty}N\cdot{\rm \widetilde{CRB}_{xyz}}\TwoOneColumnAlternate{\nonumber\\}{;\qquad}
	{\rm \widetilde{LCRB}_{xy}}\TwoOneColumnAlternate{&=&}{=}\lim_{N\rightarrow\infty}N\cdot{\rm \widetilde{CRB}_{xy}}\TwoOneColumnAlternate{\nonumber\\}{;\qquad}
	{\rm \widetilde{LCRB}_{z}}\TwoOneColumnAlternate{&=&}{=}\lim_{N\rightarrow\infty}N\cdot{\rm \widetilde{CRB}_{z}}.
	\label{lcrb_til_def}
\end{IEEEeqnarray}
The following theorem provides closed-form expressions for the latter terms.  
\begin{theorem}
	For the system defined in Sec. \ref{SEC:PORBLEMFORMULATION}, ${\rm \widetilde{LCRB}_{\rm xy}}$, ${\rm \widetilde{LCRB}_{\rm z}}$ (see \eqref{lcrb_til_def}) are given by 
	\begin{IEEEeqnarray}{rCl}
	&&\hspace{-5mm}{\rm \widetilde{LCRB}_ {xy}}=\Bigg[\eta\rho \Big(\log\left(\frac{D_{\max}}{h}\right)\frac{((r+h)^2+r^2)}{8(r+h)r^3}\TwoOneColumnAlternate{\nonumber\\
	&&}{}-\frac{r+h-r\zeta^2-\zeta D_{\max}}{8(r+h)r^2}\Big)\Bigg]^{-1}		\label{mcrb2d}
	\end{IEEEeqnarray}
	\begin{IEEEeqnarray}{rCl}
	&&\hspace{-5mm}{\rm \widetilde{LCRB}_ z}=\Bigg[\eta\rho\Bigg(\frac{r+h-r\zeta^2-\zeta D_{\max}}{2(r+h)r^2}\TwoOneColumnAlternate{\nonumber\\
	&&}{}-\frac{h(2r+h)}{2r^3(r+h)} \log\Big(\frac{D_{\max}}{h}\Big)\nonumber\\
	&&-\frac{[r+h-D_{\max}^{-1}(r(h-D_{\max}\zeta)+h(r+h))]^2}{r^3(r+h)\log\Big(\sfrac{D_{\max}}{h}\Big)}\Bigg)\Bigg]^{-1}.
	\label{mcrbz}
	\end{IEEEeqnarray}
\label{THEOREM:LCRB_TDOA}
\end{theorem}
\begin{IEEEproof}
	see Appendix \ref{PRTHEOREM:LCRB_TDOA}.
\end{IEEEproof}

Although very popular \cite{hamon1974spectral,hannan1973estimating,knapp1976generalized,yang2011efficient,yang2009efficient,myref12, myref11, weiss2003accuracy, Shang2009Multiuser, nlos1tdoa,van2007bayesian}, estimation based solely on TDOA is not optimal. In practice, however, TDOA is significantly more informative than RSS. The following lemma guarantees that RSS information is negligible for large enough bandwidth.
\begin{lemma}
	For the system defined in Sec. \ref{SEC:PORBLEMFORMULATION}, 
	\begin{IEEEeqnarray}{rCl}
		{\rm CRB_{xy}}&=&{\rm\widetilde{CRB}_{xy}}+o(W_e^{-2})\label{Lem_xy}\\
		{\rm CRB_{z}}&=&{\rm\widetilde{CRB}_{z}}+o(W_e^{-2})\label{Lem_z}.
	\end{IEEEeqnarray}	\label{LEM:CONDITION} 
\end{lemma}
\begin{IEEEproof}
	see Appendix \ref{PRLEM:CONDITION}.
\end{IEEEproof}
Furthermore, the $\rm \widetilde{CRB}$ converges to $\rm CRB$ when $K_{i}$ approaches $L_{i}$. The latter occures if
$1\ll \frac{D_i^2W_e^2}{c^2}$ for every $D_i$. As $D_i\ge h$, we conclude that RSS information is negligible for
\begin{equation}
	W_e\gg\frac{c}{h}.		\label{condition}
\end{equation}
To illustrate the latter approximation, consider a satellite at 20,000 km height. The minimal bandwidth at which (\ref{condition}) is satisfied is 15 Hz, where the typical bandwidth is at least one MHz. Thus, in most practical scenarios, the RSS effect on the estimation accuracy is negligible.

Now that we showed that estimation solely utilizing TDOA is near-optimal in most practical scenarios, we turn our attention to evaluating its performance. The latter analysis is even further motivated because TDOA localization is simpler than localization that also combines RSS.\footnote{Hence RSS is rare in practical satellite localization.} 
Moreover, the asymptotic expression for TDOA localization-accuracy, ($\rm \widetilde{LCRB}$), is much simpler than the asymptotic CRB (although nearly identical in practice).
\par 
For a sufficient number of satellites, Theorem  \ref{THEOREM:LCRB_TDOA}  provides a good approximation for the localization accuracy, $\rm \widetilde{CRB}$ (cf. \eqref{lcrb_til_def}), as follows:
\begin{IEEEeqnarray}{rCl}
	{\rm \widetilde{CRB}_{xy}}(\xivec_1^N)\TwoOneColumnAlternate{&\approx&}{\approx} {\rm \widetilde{ACRB}_{xy}}\triangleq \frac{{\rm \widetilde{LCRB}_{xy}}}{N}\TwoOneColumnAlternate{\nonumber\\}{;\qquad}
	{\rm \widetilde{CRB}_{z}}(\xivec_1^N)\TwoOneColumnAlternate{&\approx&}{\approx} {\rm \widetilde{ACRB}_{z}}\triangleq \frac{{\rm \widetilde{LCRB}_{z}}}{N}
	\label{acrbDef}
\end{IEEEeqnarray}
and combining the latter two yields the following approximated for the actual localization accuracy (see \eqref{til_CRBparts}).
Unlike the exact $\rm \widetilde{CRB}$, $\rm \widetilde{ACRB}$ is not a function of a particular  satellite configuration, but only of some general system parameters. Therefore, the latter can even be used for determining these parameters and enhanced performance already in the design phase before deploying such a massive constellation. More importantly, the exact $\rm \widetilde{CRB}$ depends on the location of each of the satellites, which continually change, while the $\rm \widetilde{ACRB}$ provides a good approximation, without considering specific satellite locations. We further demonstrate how useful $\rm \widetilde{ACRB}$ is in the next section. 
\par

%% file: Chapters/sat_numerical_results.tex
We now turn to numerical analysis for further insights and to demonstrate the theory. In all simulations, unless stated otherwise, the satellites height is $h=20000$ Km; $\eta\rho=6.4\cdot10^{13}$; the LT coverage (maximal viewing) angle is  $\phi_{\rm L,max}=60^{\circ}$ (which is equivalent to $\phi_{\rm E,max}=47.93^{\circ}$, measured from Earth's center, for satellites at $h=20000$ Km); and Earth's radius is $r=6371$ Km. Finally, as described in Sec. \ref{SEC:PORBLEMFORMULATION}, the satellites are uniformly distributed around the Earth. 
\subsection{Performance of ML Estimator}
As a first step, we show that the (non asymptotic) CRB, $\rm\widetilde{CRB}$ 
indeed predicts the performance of the TDOA based ML estimator. We consider the localization when one satellite is located precisely above the terminal and five more  satellites are evenly distributed on the circle at elevation $\phi_{\rm L}=30^{\circ}$. Hence, the received energy from the central satellite is, by definition, $E_{s,max}$, while the received energy from the other satellites is lower due to their larger distance. Following our TDOA localization approach (see Sec. \ref{chap:derivation_tdoa}), the terminal estimated the amplitudes and location jointly via the ML estimator based on (\ref{tdoamodel}). We evaluated the localization MSE by averaging the error using 1000 Monte Carlo trials and compared it to the corresponding CRB, $\rm\widetilde{CRB}$, as a function of $E_{s,max}/N_0$. \par
\begin{figure}[t!]
	\centering
	\includegraphics[width=\TwoOneColumnAlternate{0.9}{0.7}\columnwidth]{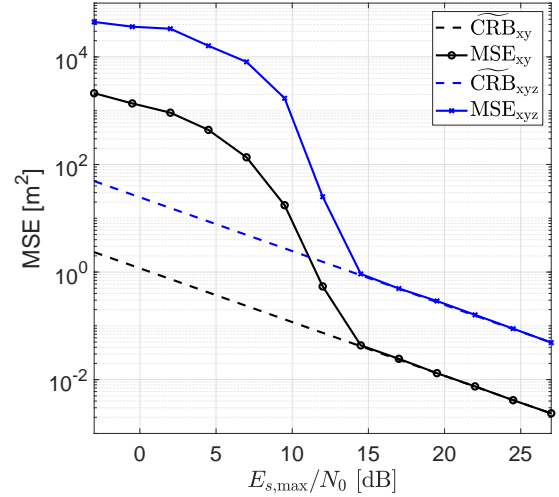}
	\caption{MSE of the TDOA receiver localization  as a function of $E_{s,\max}/N_0$, compared to the corresponding $\rm\widetilde{CRB}$. Two models are considered separately: in the first one the $z$ coordinate is considered known so only $\rm MSE_{xy}$ is examined, whereas in the second model the  $z$ coordinate is assumed to be unknown and  $\rm MSE_{xyz}$ is examined.}
	\label{fig:ml} 
\end{figure}

 
Fig. \ref{fig:ml} depicts the localization error for two estimation scenarios. In the first, we assumed that the receiver knows the $z$ coordinate, whereas, in the second, it does not, hence performing 2D and  3D TDOA-based localization, respectively. The figure presents the MSE compared to  $\rm \widetilde{CRB}_{xy}$ and $\rm \widetilde{CRB}_{xyz}$ (see (\ref{acrbDef}) and (\ref{til_CRBparts})), for the first and the second case, respectively. 
The results corroborate the well-known asymptotic efficiency of the ML estimator; i.e., beyond a threshold SNR, the estimator operates in the {\em small error regime} where the MSE approaches the lower bound. Moreover, as expected, 2D localization is more accurate than 3D localization. However, the factor of more than ten between the two indicates that estimating altitude ($z$ coordinate) is challenging compared to horizontal localization.
  
\subsection{Performance Convergence}
\begin{figure}[t]
	\centering
	\includegraphics[width=\TwoOneColumnAlternate{0.9}{0.7}\columnwidth]{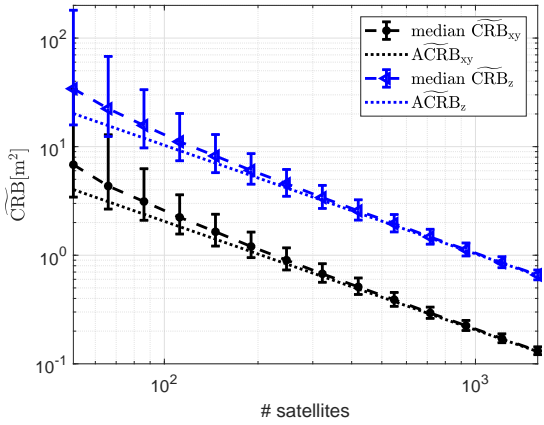}
	\caption{Empirical median of ${\rm\widetilde{CRB}}_{\rm xy}$ and ${\rm\widetilde{CRB}}_{\rm z}$ and the respective limit values as a function of $N$, the number of satellites around the Earth.}
	\label{fig:crb_convergence_M} 
\end{figure}

We now study the convergence of $N\cdot \widetilde{\rm CRB}$ to the corresponding bound $\rm \widetilde{LCRB}$. Fig. \ref{fig:hist_convs} depicts the empirical probability density function (PDF) of $N\cdot {\rm\widetilde{CRB}}$s based on 40000 random network realizations for various satellite numbers. The figure shows both the PDFs of $N\cdot{\rm\widetilde{CRB}_{xy}}$ and of $N\cdot{\rm\widetilde{CRB}_z}$, as well as the theoretical limits (\ref{mcrb2d}) and (\ref{mcrbz}). The results demonstrate the increasing accuracy of the limiting expression as the number of satellites grows. \\
\begin{figure}[t]
	\centering
	\includegraphics[width=\TwoOneColumnAlternate{0.9}{0.7}\columnwidth]{tot_hist_conv} 
	\caption{PDF of $N\cdot {\rm\widetilde{CRB}_{xy}}$ and $N\cdot {\rm\widetilde{CRB}_{z}}$ for various numbers of satellites. The figure also shows the limit values, ${\rm\widetilde{LCRB}_{xy}}$ and ${\rm\widetilde{LCRB}_{z}}$ ($\xivec_1^N$ was randomally generated and ${\rm\widetilde{CRB}}$ was calculated using $\xivec_1^N$).}
	\label{fig:hist_convs}
\end{figure}
Fig. \ref{fig:crb_convergence_M} studies the latter convergence in terms of the empirical median of the random $\rm\widetilde{CRB}$. The figure error bars (here and for all subsequent figures) mark the $10$th and $90$th percentiles of the $\rm\widetilde{CRB}$. For each simulation point, we used $200$ Monte Carlo trials, each with an independent realization of $\xivec_1^N$, from which we calculated ${\rm\widetilde{CRB}}(\xivec_1^N)$.
The figure shows that $\rm\widetilde{CRB}$ empirical median converges to the corresponding $\rm\widetilde{ACRB}$ as the number of satellites gets large. Recall that the strip created by the error bars encapsulates $80$\% of the $\rm\widetilde{CRB}$ realizations. The fact that this gap is tiny for large satellite numbers demonstrates performance convergence. Furthermore, the approximation is quite good even for a relatively small number of satellites. For example, for $420$ satellites, $80$\% of the evaluated $\rm\widetilde{CRB}_z$ were within $50$\% of the median and $80$\% of the evaluated $\rm\widetilde{CRB}_{xy}$ were within $36$\% of the median.

\subsection{Height and Angle Effect and Coverage Probability} 
\begin{figure}[t]
	\centering
	\includegraphics[width=\TwoOneColumnAlternate{0.9}{0.7}\columnwidth]{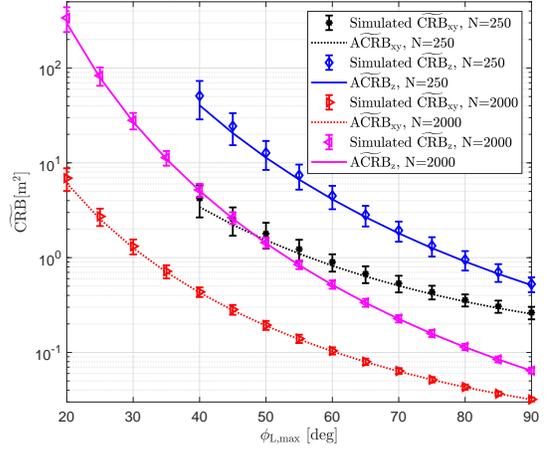}
	\caption{Empirical median of ${\rm\widetilde{CRB}}_{\rm xy}$ and ${\rm\widetilde{CRB}}_{\rm z}$ and the respective limit values $\rm\widetilde{ACRB}_{xy}$, $\rm\widetilde{ACRB}_{z}$ as a function of $\phi_{\rm L,max}$ for $N=250$ and $N=2000$ satellites around Earth.}  
	\label{fig:tot_convergence_phimax_crbe} 
\end{figure}

 We now study the effect of $h$ and $\phi_{\max}$ on the convergence rate of $\rm\widetilde{CRB}$ toward $\rm\widetilde{ACRB}$. Fig. \ref{fig:tot_convergence_phimax_crbe} depicts the median $\rm\widetilde{CRB}$ and the $\rm\widetilde{ACRB}$ as a function of the LT coverage angle, $\phi_{\rm L,max}$, 
for $250$ and $2000$ satellites uniformly distributed around Earth.
In all cases, increasing $\phi_{\rm L,max}$ reduces the ${\rm\widetilde{CRB}}$, implying improved estimation accuracy. The improvement is due to the increased LT coverage-angle that allows reception from more satellites. Moreover, the additional received satellites as $\phi_{\rm L,max}$ increases further decreases the gap between the error bars. Interestingly, the increase in $\phi_{\rm L,max}$ affects $\rm\widetilde{CRB}_z$ more than $\rm\widetilde{CRB}_{xy}$. The intuition behind this observation is that for small $\phi_{\rm L,max}$, all satellites are approximately above the terminal, thus enhancing the horizontal resolution more than the vertical one. As $\phi_{\rm L,max}$ increases, the receiver gets signals from different directions, thus improving its z-axis resolution. \par 
Note that 3D TDOA localization requires at least four satellites. Consequently, besides localization-accuracy, we need to consider also the localization-identifiability; i.e., the probability of having four satellites in the LT coverage cup. To better understand the identifiability, we define the coverage probability as the probability of having four or more satellites in the LT coverage cup. The probability that a given satellite is in the coverage cup is 
\begin{eqnarray}
p=\frac{1-\cos\phi_{\rm E,max}}{2}=\frac{h-D_{\max}\cos\phi_{\rm L,max}}{2R}.
\end{eqnarray}
Recalling that the satellite locations are i.i.d., the probability for more than three satellites in the coverage cup is 
\begin{eqnarray}
P_{\rm cov}=1-\sum_{m=0}^3 \binom{N}{m} p^m(1-p)^{N-m}.
\label{pcov}
\end{eqnarray}
\par
Note that actual satellite constellation (e.g., GPS) may achieve better coverage for the same number of satellites, as explained in Footnote $1$. This is achieved by careful planning of the satellites' orbits. However, the planning becomes very difficult for larger constellations, and does not hold at all for multi-GNSS localization. 
\par 

\begin{figure}[t]
	\centering
	\includegraphics[width=\TwoOneColumnAlternate{0.9}{0.7}\columnwidth]{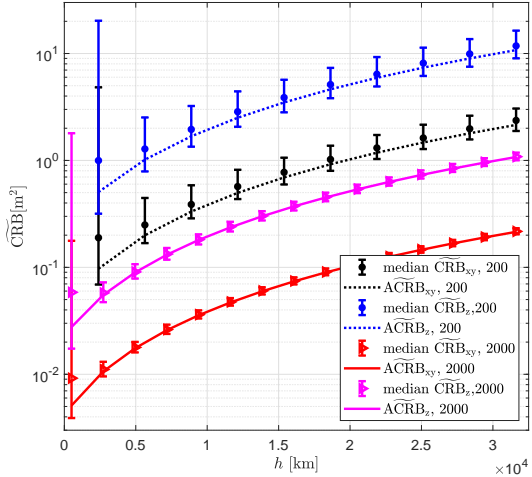}
	\caption{Empirical median of $N\cdot{\rm\widetilde{CRB}}_{\rm xy}$ and $N\cdot{\rm CRB}_{\rm z}$ and the respective ACRBs as a function of $h$. In this simulation, $N=250$.}
	\label{fig:tot_convergence_h} 
\end{figure}

From \eqref{pcov} one may evaluate the minimal coverage-angle guaranteeing a given coverage probability. For example, for satellites at 20 000 km height, a $90\%$ coverage probability is achieved with $\phi_{\rm L,max}=24.5^{\circ}$ and $\phi_{\rm L,max}=8.7^{\circ}$ for 250 and 2000 satellites, respectively. Fig. \ref{fig:tot_convergence_phimax_crbe} (as well as Fig. \ref{fig:crb_convergence_M} and Fig. \ref{fig:tot_convergence_h}) only presents results for scenarios with at least $90\%$ coverage. As the coverage probability is higher in the 2000-satellite deployment, this curve indeed starts at a smaller angle than the 250-satellite curve. 
\par 
Fig. \ref{fig:tot_convergence_h} depicts the empirical median of $\rm \widetilde{CRB}$ and the corresponding limit, ${\rm\widetilde{ACRB}}$, as a function of the satellites height, $h$. The figure considers $200$ and $2000$ satellites randomly scattered around Earth.
Note that the number of satellites within the LT coverage cup increases as $h$ increases (keeping $\phi_{\rm L,max}=60^{\circ}$), but at a lower rate than in the previous figure; i.e., changing $h$ from  5000 km to 35000 km increases the number of visible satellites by only $70\%$. Thus, the gap between the error bars decreases but does not disappear as $h$ increases. 
\par
Despite the increasing number of visible satellites, the theoretical ${\rm\widetilde{ACRB}}$ gets larger with $h$.  We thus conclude that if there is a sufficient amount of satellites, 
they should be located as closer to Earth as possible. \par
However, as Fig. \ref{fig:tot_convergence_h} shows, one should caution when using the latter conclusion with a finite number of satellites. If the satellites are too low, their average number within the LT coverage cup is small, thus rendering the asymptotic analysis unsuitable. We further discuss this trade-off in the following subsection. 
\par 

From \eqref{pcov}, the minimal height ensuring a $90\%$ coverage (which is the curves initial heights) for $\phi_{\rm L,max}=60^{\circ}$ is 2400 km and 500 km for 200 and 2000 satellites, respectively.
We, therefore, conclude that the monotonic increase of $\widetilde{\rm ACRB}$ with the height is useful only above a height that guarantees sufficient coverage.

\subsection{Insights Via Large and Small $h$ Approximations.}
We begin with characterizing $\widetilde{\rm ACRB}$ for very small and very large heights.
\begin{theorem}
	The asymptotic behavior of ${\rm \widetilde{ACRB}_{xy}}$ and ${\rm \widetilde{ACRB}_{z}}$ is described by the following limits:
	\begin{IEEEeqnarray}{rCl}
	\alpha_{\rm xy}&\triangleq& \lim_{h\rightarrow 0}\widetilde{\rm ACRB}_{\rm xy}\TwoOneColumnAlternate{\nonumber\\
	&=&}{=}\frac{-8r^2}{\eta\rho N \Big(2\log\cos\phi_{\rm L,max}+\sin^2\phi_{\rm L,max}\Big)}\label{limhACRBExy}\\
	\alpha_{\rm z}&\triangleq& \lim_{h\rightarrow 0}\widetilde{\rm ACRB}_{\rm z}\TwoOneColumnAlternate{\nonumber\\
	&=&}{=}\frac{2r^2}{\eta\rho N\Big(\sin^2\phi_{\rm L,max}+\frac{2(1-\cos\phi_{\rm L,max})^2}{\log\cos\phi_{\rm L,max}}\Big)}\label{limhACRBEz}\\
	\beta_{\rm xy}&\triangleq&\lim_{h\rightarrow \infty}\frac{{\rm \widetilde{ACRB}_{xy}}}{h^2}\TwoOneColumnAlternate{\nonumber\\
	&=&}{=}\frac{12}{\eta\rho N(\cos\phi_{\rm L,max}+2) (1-\cos\phi_{\rm L,max})^2}\label{limhACRBExyinf}\\
	\beta_{\rm z}&\triangleq&\lim_{h\rightarrow  \infty}\frac{{\rm \widetilde{ACRB}_{z}}}{h^2}=\frac{12}{\eta\rho N(1-\cos\phi_{\rm L,max})^3}\label{limhACRBEzinf}	
	\end{IEEEeqnarray}
	\label{THE:ACRB_H0_HINF}
\end{theorem}
\begin{IEEEproof}
	see Appendix \ref{PRTHE:ACRB_H0_HINF}.
\end{IEEEproof}

\begin{figure}[t]
	\centering
	\includegraphics[width=\TwoOneColumnAlternate{0.9}{0.7}\columnwidth]{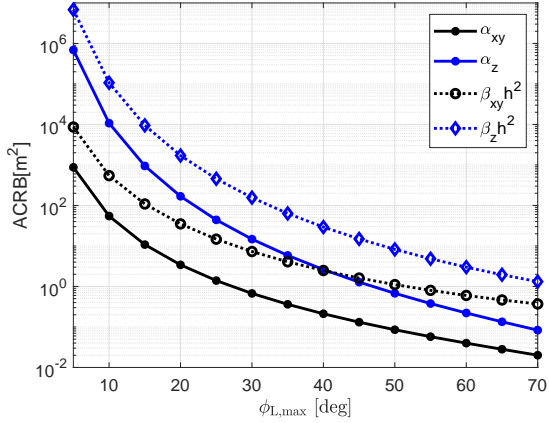}
	\caption{Two parts of ${\rm AACRB}$ produced in Theorem \ref{THE:ACRB_H0_HINF}, as a function of $\phi_{\rm L,max}$. $N=200$. The $\rm AACRB_{xy}$ scales as $\phi^{-4}$ whereas the $\rm AACRB_z$ scales as $\phi^{-6}$.}
	\label{fig:ACRB_LIMITS_phimax_h2} 
\end{figure}

Fig. \ref{fig:ACRB_LIMITS_phimax_h2} depicts the asymptotic factors of Theorem \ref{THE:ACRB_H0_HINF}. We depict each $\alpha_{xy}$ and $\alpha_z$ as is, whereas, to adjust the graph units, we multiply $\beta_{xy}$ and $\beta_z$ by $h^2$.
As expected, both $\alpha$ and $\beta$ decrease when the angle increases. Interestingly, at low angles, the $\alpha_{xy}$ and $\beta_{xy}$ scale as $\phi^{-4}$ whereas the $\alpha_{z}$ and $\beta_{z}$ scale as $\phi^{-6}$. Beyond corroboration, the aforementioned observation quantifies our previous conclusion regarding the higher dependence of the vertical estimation on coverage angle.
\par
The $h^2$ scaling of the ACRB calls for comment. Three factors affect the localization accuracy as the satellites' height increases. First, the received power from each satellite decreases as $1/h^2$. Because the ACRB scales as $1/E_s$, the effect of the satellite height matches the ACRB scaling exactly. The second factor is the increase in the number of observable satellites as $h$ increases, which improves the accuracy. The third factor is a less adequate system topology (since all the transmitters are located farther away from the target), which degrades localization accuracy. The fact that the asymptotic performance scales as $h^2$ shows that the received energy is the dominant factor, whereas the two other factors become negligible for large heights.
\par
Combining the limits from Theorem \ref{THE:ACRB_H0_HINF}, we can also produce an approximated formula, which is close to ${\rm \widetilde{ACRB}}$ but much simpler:
\begin{eqnarray}
{\rm AACRB_{xy}}&=&\alpha_{\rm xy}+\beta_{\rm xy}h^2\label{aacrbxy}\\
{\rm AACRB_{z}}&=&\alpha_{\rm z}+\beta_{\rm z}h^2
\label{aacrbz}
\end{eqnarray}

\begin{figure}[t]
	\centering
	\includegraphics[width=\TwoOneColumnAlternate{0.9}{0.7}\columnwidth]{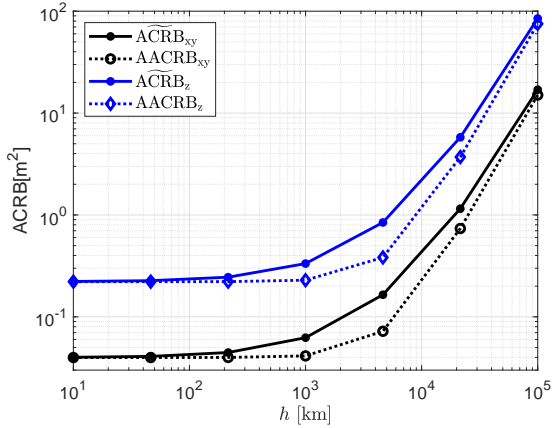}
	\caption{${\rm\widetilde{ACRB}}$ and ${\rm AACRB}$ as a function of $h$ for ${xy}$ and ${z}$ estimation.}
	\label{fig:ACRB_AACRB_h} 
\end{figure}

Fig. \ref{fig:ACRB_AACRB_h} depicts the  ${\rm\widetilde{ACRB}}$, (\ref{acrbDef}), and ${\rm AACRB}$, (\ref{aacrbxy}) and (\ref{aacrbz}), as a function of $h$, 
 corroborating the limits.  Moreover, this simple expression is a reasonable approximation for all heights. The approximation maximum deviation from the actual ${\rm\widetilde{ACRB}}$ is by a factor of $2.2$. 

%% file: Chapters/sat_conclusions.tex
We presented a new approach for analyzing TDOA and TDOA-RSS-based receiver localization via a satellite network. Assuming uniform distributed satellites on a sphere around the Earth, we used the Law of Large Numbers to calculate the localization-accuracy limiting-behavior. The resulting expressions depend solely on the network statistics and are not a function of a particular satellites locations. 
We also presented a simple condition proving that the RSS information is negligible in practical scenarios, and that the TDOA accuracy approaches the RSS+TDOA accuracy.\par
Numerical results demonstrated the convergence of the actual performance to the theoretical
prediction. We used these findings to study the performance expressions and to draw
novel insights. In particular, we showed that the ACRB increases monotonically with the height
and scales as $h^{2}$ for large $h$. Thus, as long as we guarantee a sufficient coverage probability, a lower satellite-height will result in better localization accuracy for any
number of satellites.
We also showed that the vertical localization accuracy is typically lower than the horizontal
accuracy and also more sensitive to the maximal receiver antenna angle, $\phi_{\rm L,max}$; i.e., the vertical accuracy degrades much faster than the horizontal as the  $\phi_{\rm L,max}$ decreases.
\par
While the satellites are continually moving, the derived expressions depend on the network statistics and not on the specific locations. Thus, the presented closed-form-performance expressions are convenient and straightforward and can be used for a performance approximation that does not change over time. The derived expressions also provide insights on various network parameters (e.g., number of satellites, height, power), even if the actual deployment is yet unknown. Further research is necessary to better characterize the convergence-rate and its dependency on the different parameters. 

%% file: Chapters/sat_proofs.tex
\section{Proof of Theorem \ref{THEOREM:LCRB_RSS}}\label{PRTHEOREM:LCRB_RSS}

	Using the definition of $\rm CRB_{xy}$ and $\rm CRB_z$, and the continuity of $(\cdot)^{-1}$ for matrices whose determinant is bounded away from zero:
	\begin{IEEEeqnarray}{rCl}
		{\rm LCRB_{xy}}\TwoOneColumnAlternate{&=&}{=}\left[\lim_{N\rightarrow\infty}\frac{\Jmat}{N}\right]^{-1}_{1,1}+\left[\lim_{N\rightarrow\infty}\frac{\Jmat}{N}\right]^{-1}_{2,2}\TwoOneColumnAlternate{\nonumber\\}{;\qquad}
		{\rm LCRB_{z}}\TwoOneColumnAlternate{&=&}{=}\left[\lim_{N\rightarrow\infty}\frac{\Jmat}{N}\right]^{-1}_{3,3}.
	\end{IEEEeqnarray}
	Considering the uniform random satellite locations, each of $\Jmat$ entries is the sum of i.i.d. random variables. We thus employ the Law of Large Numbers to evaluate the expectation with respect to these locations. Explicitly, representing an entry of $\Jmat$ by
	 $[\Jmat]_{m,n}=\sum_ib_i$, we use $[\lim_{N\rightarrow\infty}\frac{\Jmat}{N}]_{m,n}={\rm E}[b_i]$.
	
	Employing the Law of Total Expectation, we first condition on all elevation angles, $\{\phi_{\mathrm{L},i}\}$, and then take the expectation with respect to these angles. Accordingly, the expectation of sums involving $\sin\theta$ or $\cos\theta$ is zero, whereas for sums with $\sin^2\theta$ or $\cos^2\theta$, it equals $\frac{1}{2}$. It follows that \par 
{ \begin{IEEEeqnarray}{rCl}
&&\hspace{-7mm}E\Big[\frac{[\Jmat]_{m,n}}{N}|\{\phi_{{\rm L},i}\}\Big]\TwoOneColumnAlternate{\nonumber\\
&&}{}=\begin{cases}
	\frac{1}{2}{\rm E}[ K_i\sin^2\phi_{{\rm \tiny L},i}\mathbf{1}_{\phi_{{\rm \tiny L},i}}] & m=1,2,n=m \\
	{\rm E}[ K_i\cos^2\phi_{{\rm \tiny L},i}\mathbf{1}_{\phi_{{\rm \tiny L},i}}] & m=3,n=3 \\
	{\rm E}[L_i\cos\phi_{{\rm \tiny L},i}\mathbf{1}_{\phi_{{\rm \tiny L},i}}] & m=3,4,n=7-m \\
	{\rm E}[ L_i\phi_{{\rm \tiny L},i}\mathbf{1}_{\phi_{{\rm \tiny L},i}}] & m=4,n=4 \\
	0 & \text{o.w.}
\end{cases}
\label{EJc}
	\end{IEEEeqnarray}}
	Because ${\rm E}[\Jmat/N]$ is a block diagonal matrix, the block corresponding to $(x,y)$ in $N\cdot{\rm CRB}$ is the same as though $(z,T_{0})$ were known; implying no coupling between $(x,y)$ and $ (z,T_{0})$.  From an estimation point of view, it means that knowing the unknowing $(z,T_{0})$ does not affect the estimation accuracy of $(x,y)$ under the regime in which CRB is asymptotically tight. 
	We now calculate ${\rm LCRB_{xy}}$. From (\ref{EJc}) it follows that
	\begin{equation}
	{\rm LCRB_{xy}}=\frac{4}{{\rm E}[K_i\sin^2\phi_{{\rm \tiny L},i}\mathbf{1}_{\phi_{{\rm \tiny L},i}}]}.
	\label{LCRB2D_FIRST_c}
	\end{equation}
	Using \eqref{coordinateTransformation} and \eqref{d_i_form} one obtains
	\begin{IEEEeqnarray}{rCl}
	K_i\sin^2\phi_{{\rm \tiny L},i}\TwoOneColumnAlternate{&=&}{=}\frac{2\rho}{D_i^{4}}\Big(1+\eta D_i^2\Big)\sin^2\phi_{{\rm \tiny L},i}\TwoOneColumnAlternate{\nonumber\\
	&=&}{=}2\rho R^2D_i^{-6}\Big(1+\eta D_i^2\Big)\sin^2\phi_{{\rm E},i}.
	\end{IEEEeqnarray}
	Noting that $\chi\triangleq \cos\phi_{{\rm E},i}$ is uniformly distributed over $[-1,1]$, and defining $\chi_{\max}\triangleq\cos\phi_{\rm E,max}$, one obtains:
	\begin{IEEEeqnarray}{rCl}
	\TwoOneColumnAlternate{&&\hspace{-7mm}}{}{\rm E}[K_i\sin^2\phi_{{\rm \tiny L},i}\mathbf{1}_{\phi_{{\rm \tiny L},i}}]\TwoOneColumnAlternate{=}{&=&}\rho R^2\int_{-1}^{1}\frac{1-\chi^2}{(R^2+r^2-2rR\chi)^3}\TwoOneColumnAlternate{\nonumber\\
	&&\cdot}{}\Big(1+\eta(R^2+r^2-2rR\chi)\Big)\mathbf{1}_{\chi\in[\chi_{\max},1]}d\chi\nonumber\\
	&=&\frac{\rho}{16r^3 R}\Bigg[4 \left(2\eta\left(r^2+R^2\right)-1\right) \log \left(\frac{D_{\max}}{h}\right)\nonumber\\
	&&+\frac{(2\eta(r^2-R^2)^2-4 (r^2+R^2))(h^2-D_{\max}^2)}{D_{\max}^2h^2}\nonumber\\
	&&-4\eta r R(1-\chi_{\max})+\frac{\left(r^2-R^2\right)^2(h^4-D_{\max}^4)}{D_{\max}^4h^4}\Bigg]
	\label{EAsin2}
	\end{IEEEeqnarray}
	where $D_{\max}$ is given in (\ref{DM}). Substituting \eqref{coordinateTransformation} and simplifying establishes the desired result, (\ref{mcrb2drss}). \par
	To derive $\rm LCRB_{z}$, we consider the inverse of the second block in \eqref{EJc}, which yields
	\begin{IEEEeqnarray}{rCl}
	{\rm LCRB_{z}}&=&{\rm E}[L_i\mathbf{1}_{\phi_{{\rm \tiny L},i}}]\cdot\big({\rm E}[K_i \cos^2\phi_{{\rm \tiny L},i}\mathbf{1}_{\phi_{{\rm \tiny L},i}}]{\rm E}[L_i\mathbf{1}_{\phi_{{\rm \tiny L},i}}]\TwoOneColumnAlternate{\nonumber\\
&&}{}	-{\rm E}^2[L_i \cos\phi_{{\rm \tiny L},i}\mathbf{1}_{\phi_{{\rm \tiny L},i}}]\big)^{-1}.
	\label{LCRBZ_FIRST_c}
	\end{IEEEeqnarray}
	Considering the numerator and, once again, transforming to $\rm E$-coordinates one obtains
	\begin{IEEEeqnarray}{rCl}
	{\rm E}\Big[L_i\mathbf{1}_{\phi_{{\rm \tiny L},i}}\Big]&=&\eta\rho\int_{-1}^{1}\frac{\mathbf{1}_{\chi\in[\chi_{\max},1]}}{R^2+r^2-2rR\chi}d\chi\TwoOneColumnAlternate{\nonumber\\
	&=&}{=}\frac{\eta\rho}{rR}\log\Big(\frac{D_{\max}}{h}\Big).
	\label{EBi}
	\end{IEEEeqnarray}
	Examining the denominator of \eqref{LCRBZ_FIRST_c}, the first term is given by
	\begin{IEEEeqnarray}{rCl}
	\TwoOneColumnAlternate{&&\hspace{-7mm}}{}{\rm E}[K_i \cos^2\phi_{{\rm \tiny L},i}\mathbf{1}_{\phi_{{\rm \tiny L},i}}]\TwoOneColumnAlternate{\nonumber\\}{}
	&=&{\rm E}\left[\frac{2\rho}{D_i^{4}}\Big(1+\eta D_i^2\Big)\cos^2\phi_{{\rm \tiny L},i}\mathbf{1}_{\phi_{{\rm \tiny L},i}}\right]\nonumber\\
	&=&{\rm E}\left[\frac{2\rho}{D_i^{6}}\Big(1+\eta D_i^2\Big)(R\cos\phi_{{\rm E},i}-r)^2\mathbf{1}_{\phi_{{\rm E},i}\in[0,\phi_{\max,e}]}\right]\nonumber\\
	&=& \frac{\rho}{16r^3R}  \Bigg[\frac{2 \left(r^2-R^2\right) \left(\eta \left(r^2-R^2\right)+2\right)(D_{\max}^{2}-h^{2})}{h^2D_{\max}^{2}}\TwoOneColumnAlternate{\nonumber\\
	&&}{}-4 (2\eta  (R^2-r^2)\TwoOneColumnAlternate{}{\nonumber\\
	&&}-1) \log \left(\frac{D_{\max}}{h}\right)\TwoOneColumnAlternate{\nonumber\\
	&&}{}+4 \eta rR (1-\chi_{\max})+\frac{\left(r^2-R^2\right)^2(D_{\max}^{4}-h^{4})}{h^{4}D_{\max}^{4}}\Bigg].
	\end{IEEEeqnarray}
	Now to ${\rm E}[L_i\cos\phi_{{\rm \tiny L},i}\mathbf{1}_{\phi_{{\rm \tiny L},i}}]$, using \eqref{coordinateTransformation} again:
	\begin{IEEEeqnarray}{rCl}
	{\rm E}[L_i\cos\phi_{{\rm \tiny L},i}\mathbf{1}_{\phi_{{\rm \tiny L},i}}]&=&{\rm E}\Big[\frac{2\eta\rho}{D_i^{3}}(R\cos\phi_{{\rm E},i}-r)\mathbf{1}_{\phi_{{\rm E},i}}\Big]\TwoOneColumnAlternate{\nonumber\\
	&=&}{=}\frac{\eta\rho}{r^2}\Big(1-\frac{R-r\chi_{\max}}{D_{\max}}\Big).
	\end{IEEEeqnarray}
	Combining all together and simplifying yields \eqref{mcrbzrss}. 

\section{Decoupling of {$\avec$} and $\gammavec$}	\label{chap:appendixA}
This appendix provides a proof for the following theorem:
\begin{theorem}
	Assuming the signal model \eqref{tdoamodel}, there is no coupling between the $\gammavec$ estimation and the ${\bf a}$ estimation. 
\end{theorem}
\begin{IEEEproof}
	To prove the theorem we show that FIM cross-entries are zero. That is, we need to prove that
	 \begin{IEEEeqnarray}{rCl}
	 	&&\hspace{-7mm}{\rm E}\left[\frac{\partial^2 L(\tilde{\gamma})}{\partial A_m\partial x}|\tilde{\gamma}\right]={\rm E}\left[\frac{\partial^2 L(\tilde{\gamma})}{\partial A_m\partial y}|\tilde{\gamma}\right]={\rm E}\left[\frac{\partial^2 L(\tilde{\gamma})}{\partial A_m\partial z}|\tilde{\gamma}\right]\TwoOneColumnAlternate{\nonumber\\
	 	&=&}{=}{\rm E}\left[\frac{\partial^2 L(\tilde{\gamma})}{\partial A_m\partial T_0}|\tilde{\gamma}\right]=0, \quad\forall m.
	 \end{IEEEeqnarray} 
	 The log likelihood function is given (up to an additive constant) by
	\begin{IEEEeqnarray}{rCl}
	L(\tilde{\gamma})&=&\sum_{m=1}^{M}\log(f(\tilde{r}_{m}|\tilde{\gamma}))\TwoOneColumnAlternate{\\
	&=&}{=}-\sum_{m=1}^{M} \frac{1}{N_0}\int_{0}^{\infty}(\tilde{r}_{m}(t)-A_ms(t-\tau_m(\tilde{\gamma})))^2dt\TwoOneColumnAlternate{\nonumber}{}
	\label{Ltilgama}
	\end{IEEEeqnarray}
	where $\tau_m(\tilde{\gamma})=\frac{D_m(\tilde{\gamma})}{c}+T_0$. To obtain the cross entries, we differentiate \eqref{Ltilgama} with respect to (w.r.t.) $A_{m}$ and then w.r.t. the location or the time coordinate. The log likelihood derivative w.r.t. $A_m$ is
	\begin{IEEEeqnarray}{rCl}
	\frac{\partial L(\tilde{\gamma})}{\partial A_m}&=&\frac{2}{N_0}\int_{0}^{\infty}s(t-\tau_m(\tilde{\gamma}))\tilde{r}_{m}(t)dt\TwoOneColumnAlternate{\nonumber\\
	&&}{}-\frac{2}{N_0}\int_{0}^{\infty}A_ms^2(t-\tau_m(\tilde{\gamma}))dt.
	\label{firstder}
	\end{IEEEeqnarray}
	\par 
	Now, examining the cross entries corresponding to $T_0$ and $A_m$, the derivative of \eqref{firstder} w.r.t. $T_0$ is
	\begin{IEEEeqnarray}{rCl}
		\TwoOneColumnAlternate{&&\hspace{-7mm}}{}\frac{\partial^2 L(\tilde{\gamma})}{\partial A_m\partial T_0}\TwoOneColumnAlternate{=}{&=&}-\frac{2}{N_0}\int_{0}^{\infty}\dot{s}(t-\tau_m(\tilde{\gamma}))\tilde{r}_{m}(t)dt\TwoOneColumnAlternate{\\
		&&}{}+\frac{4}{N_0}\int_{0}^{\infty}A_ms(t-\tau_m(\tilde{\gamma}))\dot{s}(t-\tau_m(\tilde{\gamma}))dt\nonumber\\
		&=&-\frac{2}{N_0}\int_{0}^{\infty}\dot{s}(t-\tau_m(\tilde{\gamma}))[\tilde{r}_{m}(t)-2A_ms(t-\tau_m(\tilde{\gamma}))]dt.\TwoOneColumnAlternate{\nonumber}{}
	\end{IEEEeqnarray}
	Noting that ${\rm E}[\tilde{r}_{m}(t)]=A_ms(t-\tau_m(\tilde{\gamma}))$ and
	\begin{equation}
	\int_{0}^{\infty}s(t)\dot{s}(t)dt=0
	\end{equation}
it follows that
	\begin{IEEEeqnarray}{rCl}
		{\rm E}\left[\frac{\partial^2 L(\tilde{\gamma})}{\partial A_m\partial T_0}|\tilde{\gamma}\right]&=&\frac{2A_m}{N_0}\int_{0}^{\infty}\dot{s}(t-\tau_m(\tilde{\gamma}))s(t-\tau_m(\tilde{\gamma}))dt\TwoOneColumnAlternate{\nonumber\\&=&}{=}0.
	\end{IEEEeqnarray} 
	Next, to the cross entries corresponding to $x$ and $A_m$. The parameter $x$ affects the likelihood via $D_m$. Since we use TDOA while ignoring RSS, only $\tau_m(\tilde{\gamma})$ is a function of $D_m$. Further deriving \eqref{firstder} w.r.t. $x$, we obtain \eqref{d2LdAdx}, from which \eqref{Ed2LdAdx} follows: 
\begin{IEEEeqnarray}{rCl}
	\frac{\partial^2 L(\tilde{\gamma})}{\partial A_m\partial x}&=&-\frac{2}{cN_0}\int_{0}^{\infty}\hspace{-2mm}\dot{s}(t-\tau_m(\tilde{\gamma}))\tilde{r}_{m}(t)\sin\phi_{{\rm E},m}\cos\theta_{{\rm E},m}dt\nonumber\\
	&&\TwoOneColumnAlternate{\hspace{-1.6cm}}{}+\frac{4}{cN_0}\int_{0}^{\infty}\hspace{-2mm}A_ms(t-\tau_m(\tilde{\gamma}))\dot{s}(t-\tau_m(\tilde{\gamma}))\sin\phi_{{\rm E},m}\cos\theta_{{\rm E},m}dt\nonumber\\
	&=&-\frac{2\sin\phi_{{\rm E},m}\cos\theta_{{\rm E},m}}{cN_0}\TwoOneColumnAlternate{\nonumber\\
	&&\hspace{-5mm}}{}\int_{0}^{\infty}\hspace{-2mm}\dot{s}(t-\tau_m(\tilde{\gamma}))[\tilde{r}_{m}(t)-2A_ms(t-\tau_m(\tilde{\gamma}))]dt
	\label{d2LdAdx}
\end{IEEEeqnarray}
\begin{IEEEeqnarray}{rCl}
		\label{Ed2LdAdx}
	{\rm E}\left[\frac{\partial^2 L(\tilde{\gamma})}{\partial A_m\partial x}|\tilde{\gamma}\right]&=&\frac{2\sin\phi_{{\rm E},m}\cos\theta_{{\rm E},m}}{cN_0}\TwoOneColumnAlternate{\\
	&&}{}\int_{0}^{\infty}\hspace{-2mm}\dot{s}(t-\tau_m(\tilde{\gamma}))A_ms(t-\tau_m(\tilde{\gamma}))dt=0\TwoOneColumnAlternate{\nonumber}{}
\end{IEEEeqnarray}
The proof for the cross elements corresponding to the $y$- and $z$-coordinates follow exactly the same reasoning as \eqref{d2LdAdx} and \eqref{Ed2LdAdx}.
\end{IEEEproof}
\section{Proof of Theorem \ref{THEOREM:LCRB_TDOA}}\label{PRTHEOREM:LCRB_TDOA}

Following the same approach as in Appendix A, we have that 
\begin{IEEEeqnarray}{rCl}
	{\rm\widetilde{LCRB}_{xy}} &=& {\rm E}[[\Jmat]_{11}/N]^{-1}+ {\rm E}[[\Jmat]_{22}/N]^{-1}\TwoOneColumnAlternate{\nonumber\\}{;\qquad}
	{\rm \widetilde{LCRB}_z} \TwoOneColumnAlternate{&=&}{=} {\rm E}[[\Jmat]_{33}/N]^{-1}.
\end{IEEEeqnarray}
Using again, the Law of Total Expectation, we first consider the expectation of the FIM \eqref{J3D_t} w.r.t. to $\theta$.
	
\begin{IEEEeqnarray}{rCl}
		\TwoOneColumnAlternate{&&\hspace{-7mm}}{}E\Big[\frac{[\Jmat]_{m,n}}{N}|\{\phi_{{\rm L},i}\}\Big]\TwoOneColumnAlternate{\nonumber\\
		&&}{}=\begin{cases}
			\frac{1}{2}{\rm E}[ L_i\sin^2\phi_{{\rm \tiny L},i}\mathbf{1}_{\phi_{{\rm \tiny L},i}}] & m=1,2,n=m \\
			{\rm E}[ L_i\cos^2\phi_{{\rm \tiny L},i}\mathbf{1}_{\phi_{{\rm \tiny L},i}}] & m=3,n=3 \\
			{\rm E}[L_i\cos\phi_{{\rm \tiny L},i}\mathbf{1}_{\phi_{{\rm \tiny L},i}}] & m=3,4,n=7-m \\
			{\rm E}[ L_i\phi_{{\rm \tiny L},i}\mathbf{1}_{\phi_{{\rm \tiny L},i}}] & m=4,n=4 \\
			0 & \text{o.w.}
		\end{cases}
		\label{EJ3D}
\end{IEEEeqnarray}
Beginning with ${\rm \widetilde{LCRB}}_{\rm xy}$, it follows from \eqref{EJ3D} that
	\begin{equation}
	{\rm \widetilde{LCRB}_{\rm xy}}=\frac{4}{{\rm E}[L_i\sin^2\phi_{{\rm \tiny L},i}\mathbf{1}_{\phi_{{\rm \tiny L},i}}]}
	\label{LCRB2D_FIRST}
	\end{equation}
	and as in (\ref{EAsin2}), one obtains 
	\begin{IEEEeqnarray}{rCl}
		\TwoOneColumnAlternate{&&\hspace{-7mm}}{}{\rm E}\left[L_i\sin^2\phi_{{\rm \tiny L},i}\mathbf{1}_{\phi_{{\rm \tiny L},i}}\right]\TwoOneColumnAlternate{\nonumber\\}{}
		&=& {\rm E}\left[\frac{2\rho\eta }{D_i^{2}}\left(\frac{R}{D_i}\sin\phi_{{\rm E},i}\right)^2\mathbf{1}_{\phi_{{\rm E},i}}\right]\TwoOneColumnAlternate{\nonumber\\
		&=&}{=}2\rho\eta \Big(\log\left(\frac{D_{\max}}{h}\right)\frac{((r+h)^2+r^2)}{4r^3(r+h)}\nonumber\\
		&&-\frac{(1-\chi_{\max})(D_{\max}^2+r(r+h)(1+\chi_{\max}))}{4r^2D_{\max}^2}\Big).\TwoOneColumnAlternate{\nonumber\\
		&&}{}
	\end{IEEEeqnarray}
	Finally, by transforming the polar angle back to L-coordinates using Equation \eqref{coordinateTransformation}, the desired result follows.	\par
	We now derive ${\rm \widetilde{LCRB}}_{z}$. From (\ref{EJ3D}) one obtains
	\begin{IEEEeqnarray}{rCl}
	{\rm \widetilde{LCRB}}_{\rm z}&=&{\rm E}[L_i\mathbf{1}_{\phi_{{\rm \tiny L},i}}]\cdot\big({\rm E}[L_i \cos^2\phi_{{\rm \tiny L},i}\mathbf{1}_{\phi_{{\rm \tiny L},i}}]{\rm E}[L_i]\TwoOneColumnAlternate{\nonumber\\
	&&}{}-{\rm E}^2[L_i \cos\phi_{{\rm \tiny L},i}\mathbf{1}_{\phi_{{\rm \tiny L},i}}]\big)^{-1}
	\label{MCRBz}
	\end{IEEEeqnarray}
We already evaluated all the expectations in (\ref{MCRBz}) in the proof of Theorem \ref{THEOREM:LCRB_RSS}, except of ${\rm E}[L_{i} \cos \phi_{{\rm L},i}\onevec_{\phi_{{\rm L},i}}]$. The latter is given by
	\begin{IEEEeqnarray}{rCl}
		\TwoOneColumnAlternate{&&\hspace{-7mm}}{}{\rm E}[L_i \cos\phi_{{\rm \tiny L},i}\mathbf{1}_{\phi_{{\rm \tiny L},i}}]\TwoOneColumnAlternate{\nonumber\\}{}
		&=&2\rho\eta {\rm E}[D_i^{-3}(R\cos\phi_{{\rm E},i}-r)\mathbf{1}_{\phi_{{\rm E},i}}]\TwoOneColumnAlternate{\nonumber\\
		&=&}{=}2\rho\eta \frac{1+D_{\max}^{-1}(r(\chi_{\max}-1)-h)}{2r^2}\nonumber\\
		&=&2\rho\eta \frac{D_{\max}(r+h)-r(h-D_{\max}\zeta)-(r+h)h}{2(r+h)D_{\max}r^2}\nonumber\\
		&=&2\rho\eta \frac{D_{\max}(r(1+\zeta)+h)-2rh-h^2}{2(r+h)D_{\max}r^2}
		\label{EAicos2phi}
	\end{IEEEeqnarray}
	where $\zeta=\cos\phi_{\rm L,max}$. Substituting all terms back in (\ref{MCRBz}) establishes (\ref{mcrbz}).

\section{Proof of Lemma \ref{LEM:CONDITION}}\label{PRLEM:CONDITION}
	
	Comparing (\ref{J3D_t}) to (\ref{J3Drss}), it follows that 
	\begin{eqnarray}
		{\rm J=\tilde{J}}+o(W_e).
		\label{J_oWe}
	\end{eqnarray}
	From \cite{johnson1985matrix}, Section 5.8, it follows that:
	\begin{IEEEeqnarray}{rCl}
		\|{\rm \Jmat^{-1}-\tilde{\Jmat}^{-1}}\|=o(W_e^{-2}).
	\end{IEEEeqnarray}
	which leads to \eqref{Lem_xy} and \eqref{Lem_z}. 
	
\section{Proof of Theorem \ref{THE:ACRB_H0_HINF}}\label{PRTHE:ACRB_H0_HINF}
	\subsection{The Limit of $\rm \widetilde{ ACRB}_{xy}$ as $h$ approaches $0$}

	Using a second order Taylor expansion, one obtains 
	\begin{equation}
	D_{\max}= \frac{h}{\zeta}-h^2\frac{1-\zeta^2}{2r\zeta^3}+o(h^2).
	\end{equation}
	It therefore follows that 
	\begin{equation}
		\lim_{h\rightarrow 0} \frac{D_{\max}}{h}=\frac{1}{\zeta}
		\label{limDm_H}
	\end{equation}
	and
	\begin{IEEEeqnarray}{rCl}
	h-D_{\max}\zeta&=& h^2\frac{1-\zeta^2}{2r\zeta^2}+o(h^2).
	\label{tylr}
	\end{IEEEeqnarray}
	Using some arithmetic manipulations on \eqref{mcrb2d}, we can write:
	\begin{IEEEeqnarray}{rCl}
	\TwoOneColumnAlternate{&&\hspace{-7mm}}{}{\rm \widetilde{ACRB}_{xy}}\TwoOneColumnAlternate{\nonumber\\}{}
	&=&4(r+h)\cdot\Big(\rho\eta  N \Big(\log\left(\frac{D_{\max}}{h}\right)\frac{(r+h)^2+r^2}{2r^3}\nonumber\\
	&&-\frac{h-D_{\max}\zeta}{2r^2D_{\max}^2}((r+h)^2+r(r+h-D_{\max}\zeta))\Big)\Big)^{-1}.
\IEEEeqnarraynumspace
	\end{IEEEeqnarray}
	Substituting \eqref{tylr}, one obtains:
	\begin{IEEEeqnarray}{rCl}
	\TwoOneColumnAlternate{&&\hspace{-7mm}}{}{\rm \widetilde{ACRB}_{xy}}\TwoOneColumnAlternate{\nonumber\\
	&&=}{&=&}4(r+h)\cdot\Big(\rho\eta  N \Big(\log\left(\frac{D_{\max}}{h}\right)\frac{(r+h)^2+r^2}{2r^3}\nonumber\\
	&&-\frac{h^2(1-\zeta^2)}{4r^3D_{\max}^2\zeta^2}\Big((r+h)^2+r^2+\frac{h^2(1-\zeta^2)}{2\zeta^2}\TwoOneColumnAlternate{\nonumber\\
	&&}{}+o(h^2)\Big)\Big)\Big)^{-1}.
	\IEEEeqnarraynumspace
	\end{IEEEeqnarray}
	Evaluating the limit as $h \rightarrow 0$, using the continuity of ${\rm \widetilde{ACRB}_{xy}}$ and \eqref{limDm_H}, we get
	\begin{IEEEeqnarray}{rCl}
	\alpha_{xy}&=&\lim_{h\rightarrow 0} {\rm \widetilde{ ACRB}_{xy}}\TwoOneColumnAlternate{\nonumber\\
	 &=&}{=}\frac{4r}{\rho\eta  N \Big(-\log(\zeta)\frac{r^2+r^2}{2r^3}-\frac{1-\zeta^2}{4r^3}\Big(r^2+r^2\Big)\Big)}.
	\end{IEEEeqnarray}
	Reorganizing this expression and substituting $\zeta=\cos\phi_{\rm L,max}$ completes the proof of \eqref{limhACRBExy}. 
	
	\subsection{The Limit of ${\rm \widetilde{ACRB}_{z}}$ as $h$ approaches $0$}
	From \eqref{mcrbz} and \eqref{acrbDef}, the ${\rm \widetilde{ACRB}_{z}}$ is given by:
	\begin{IEEEeqnarray}{rCl}
\TwoOneColumnAlternate{&&	\hspace{-7mm}}{}{\rm \widetilde{ACRB}_{z}}\TwoOneColumnAlternate{=}{&=&}\Bigg[\eta\rho N\Bigg(\frac{r+h-r\zeta^2-\zeta D_{\max}}{2(r+h)r^2}\TwoOneColumnAlternate{\nonumber\\
	&&}{}-\frac{h(2r+h)}{2r^3(r+h)} \log\Big(\frac{D_{\max}}{h}\Big)\nonumber\\
	&&-\frac{(r+h-D_{\max}^{-1}(r(h-D_{\max}\zeta)+h(r+h)))^2}{r^3(r+h)\log\Big(\sfrac{D_{\max}}{h}\Big)}\Bigg)\Bigg]^{-1}\TwoOneColumnAlternate{\hspace{-2mm}.}{\hspace{-4mm}.\IEEEeqnarraynumspace}\label{acrbz}
	\end{IEEEeqnarray}
	Using \eqref{tylr}, we have
	\begin{IEEEeqnarray}{rCl}
	\TwoOneColumnAlternate{&&}{}	{\rm \widetilde{ACRB}_{z}}\TwoOneColumnAlternate{=}{&=&}\Bigg[\eta\rh N\Bigg(\frac{r-r\zeta^2+h^2\frac{1-\zeta^2}{2r\zeta^2}+o(h^2)}{2(r+h)r^2}\TwoOneColumnAlternate{\nonumber\\
	&&\hspace{5mm}}{}-\frac{h(2r+h)}{2r^3(r+h)} \log\Big(\frac{D_{\max}}{h}\Big)\nonumber\\
	&&\TwoOneColumnAlternate{\hspace{5mm}}{}-\frac{(r+h-D_{\max}^{-1}(h^2\frac{1-\zeta^2}{2\zeta^2}+o(h^2)+h(r+h)))^2}{r^3(r+h)\log\Big(\sfrac{D_{\max}}{h}\Big)}\Bigg)\Bigg]^{-1}\hspace{-4mm}.
	\IEEEeqnarraynumspace
 	\end{IEEEeqnarray}
	We then substitute \eqref{limDm_H} and $h=0$ and get:
	\begin{IEEEeqnarray}{rCl}
\alpha_{z}&=&\lim_{h\rightarrow 0} {\rm \widetilde{ ACRB}_{z}}=\Bigg[\eta\rho N\Bigg(\frac{r-r\zeta^2}{2r^3}+\frac{[r-\zeta r]^2}{r^4\log(\zeta)}\Bigg)\Bigg]^{-1}
	\end{IEEEeqnarray}
	which directly leads to \eqref{limhACRBEz}.  

\subsection{The Limit of ${\rm \widetilde{ACRB}_{xy}}$ as $h$ approaches Infinity}	
	Evaluating the asymptotic behavior of $\log\left(\frac{D_{\max}}{h}\right)$ and $\sqrt{h^2+2hr+r^2\zeta^2}$ we have:
	\begin{IEEEeqnarray}{rCl}
	\TwoOneColumnAlternate{&&\hspace{-5mm}}{}\log\left(\frac{D_{\max}}{h}\right)\TwoOneColumnAlternate{=}{&=&}-\frac{r (\zeta-1)}{h}+\frac{r^2 (\zeta-1)}{h^2}\TwoOneColumnAlternate{\nonumber\\
	&&}{}+\frac{r^3 \left(\zeta^3-9\zeta+8\right)}{6 h^3}+o(h^{-3})\nonumber\\
	\TwoOneColumnAlternate{&&=}{&=&}\frac{r(\zeta-1)}{6 h^3}\left(-6 h^2+6 h r+r^2 \zeta  (\zeta+1)-8 r^2\right) \label{approxLog}\\
	\TwoOneColumnAlternate{&&\hspace{-5mm}}{}\sqrt{h^2+2 h r+r^2\zeta ^2}\TwoOneColumnAlternate{=}{&=&}\frac{-r^4 (\zeta^4-6\zeta^2+5)}{8 h^3}-\frac{r^3 \left(\zeta^2-1\right)}{2 h^2}\TwoOneColumnAlternate{\nonumber\\
	&&}{}+\frac{r^2 (\zeta^2-1)}{2 h}+h+r+o(h^{-3}).
	\label{approxSqrt}
	\end{IEEEeqnarray}
	Substituting the latter two into ${\rm \widetilde{ACRB}_{xy}}$ and reorganizing yields:
	\begin{IEEEeqnarray}{rCl}
	\TwoOneColumnAlternate{&&\hspace{-5mm}}{}{\rm \widetilde{ACRB}_{xy}}\TwoOneColumnAlternate{=}{&=&}\frac{4 r^3 (h+r)}{ \rho\eta N}\cdot\Big(\frac{r \left((h+r)^2+r^2\right) (\zeta-1)}{6 h^3}\TwoOneColumnAlternate{\nonumber\\
	&&\cdot }{}\left(-6 h^2+6 h r+r^2 \zeta  (\zeta+1)-8 r^2\right)\nonumber\\
	&&+r \zeta \Big(\frac{r^2 (\zeta^2-1)}{2 h}+h+r\Big)-r (h+r)+o(h^{-1})\Big)^{-1}\nonumber\\\TwoOneColumnAlternate{
	&&\hspace{0.9cm}=}{&=&}\frac{12 r^3h^3 (h+r)}{ \rho\eta Nh^2 r^3(1-\zeta)^2 (\zeta+2)+o(h^2)}.
	\end{IEEEeqnarray}
	Dividing the ${\rm \widetilde{ACRB}_{xy}}$ by $h^2$ and taking the limit, one obtains
	\begin{equation}
	\lim_{h\rightarrow \infty}\frac{{\rm \widetilde{ACRB}_{xy}}}{h^2}=\frac{12}{\rho\eta  N(\zeta+2) (1-\zeta)^2}
	\end{equation}
	\subsection{The Limit of ${\rm \widetilde{ACRB}_{z}}$ as $h$ approaches Infinity}
	Substituting \eqref{approxLog} and \eqref{approxSqrt} in \eqref{acrbz} and reorganizing, we get:
	\begin{IEEEeqnarray}{rCl}
\TwoOneColumnAlternate{&&	\hspace{-3mm}}{}{\rm \widetilde{ACRB}_{z}}\TwoOneColumnAlternate{=}{&=&}\Bigg[\eta\rho N\Bigg((r+h-\zeta (\frac{-r^4 (\zeta^4-6\zeta^2+5)}{8 h^3}\TwoOneColumnAlternate{\nonumber\\
&&}{}+\frac{r^2 (\zeta^2-1)}{2 h}+h+r))-\frac{r^3 \left(\zeta^2-1\right)}{2 h^2}\TwoOneColumnAlternate{}{\nonumber\\
&&}\cdot(2(r+h)r^2)^{-1}\TwoOneColumnAlternate{\nonumber\\
&&}{}+\frac{h(2r+h)}{2r^3(r+h)} \Big(\frac{6r (\zeta-1)(h^2-6 h r-r^2\left(\zeta^2+\zeta-8\right))}{6 h^3}\Big)\nonumber\\
	&&+\Big(2h^2+r\Big(r(\zeta^2-1)+\frac{r^4 (\zeta^4-6\zeta^2+5)}{4 h^3}\TwoOneColumnAlternate{\nonumber\\
	&&}{}+\frac{r^3 \left(\zeta^2-1\right)}{h^2}-\frac{r^2 (\zeta^2-1)}{ h}-2h\Big)\Big)\Big)\nonumber\\
	&&\cdot\Big((r+h)\Big(\frac{6r^4 (\zeta-1)(h^2-6 h r-r^2\left(\zeta^2+\zeta-8\right))}{6 h^3}\Big)\Big)^{-1}\nonumber\\
	&&-\Big(2h\Big(\frac{-r^4 (\zeta^4-6\zeta^2+5)}{8 h^3}-\frac{r^3 \left(\zeta^2-1\right)}{2 h^2}+\frac{r^2 (\zeta^2-1)}{2 h}\TwoOneColumnAlternate{\nonumber\\
	&&}{}+h-r\Big)\Big)\cdot\Bigg((r+h)\nonumber\\
	&&\cdot\frac{6r^4 (\zeta-1)(h^2-6 h r-r^2\left(\zeta^2+\zeta-8\right))}{6 h^3}\Bigg)^{-1}\Bigg)\Bigg]^{-1}\nonumber\TwoOneColumnAlternate{
	\end{IEEEeqnarray}
	\begin{IEEEeqnarray}{rCl}}{\\}
	&=&\Big[96 h^3 (h+r) \left(6 h^2-6 h r-\left(\zeta^2+\zeta-8\right) r^2\right)\Big]\nonumber\\
	&&\cdot \Big[\rho\eta  N (\zeta-1)\Big(4 h r^3(\zeta-1) (7 \zeta^3+30\zeta^2-15\zeta-166)\nonumber\\
	&&-48 h^3 (\zeta^2+7\zeta-14) r-48 h^4 (\zeta-1)^2\TwoOneColumnAlternate{\nonumber\\
	&&}{}+3 (\zeta+1) \left(\zeta^2-5\right) \left(\zeta^3+\zeta^2+\zeta-15\right) r^4\nonumber\\
	&&-4 h^2 (\zeta-1) (\zeta+5) (\zeta^2+7\zeta-26) r^2\Big)\Big]^{-1}+o(h^{-2}).
	\IEEEeqnarraynumspace
	\end{IEEEeqnarray}
	Dividing by $h^2$ and taking the limit, we immediately get:
	\begin{equation}
	\lim_{h\rightarrow \infty}\frac{{\rm \widetilde{ACRB}_{z}}}{h^2}=\frac{12}{\rho\eta  N(1-\zeta)^3}
	\end{equation}